\documentclass[12pt,a4paper,oneside]{article} 
\usepackage{animate}  
\usepackage[final]{pdfpages} 
\usepackage{color}  
\usepackage{enumitem} 
 
\usepackage{multirow}
\usepackage{amsmath, amsfonts, amssymb}  

\usepackage{forarray}
\renewcommand{\thefigure}{\arabic{figure}} 
\renewcommand{\thetable}{\arabic{TABLE}}
 
\usepackage{animate}

\usepackage{rotating} 
\usepackage{etoolbox}
\newtoggle{FigsAtEnd}
 
\settoggle{FigsAtEnd}{true}
\usepackage{amsmath}
\usepackage{lscape} 

\usepackage[para,online,flushleft]{threeparttable}

\usepackage{adjustbox} 
\usepackage{pifont}

\usepackage{longtable}
\usepackage{color, colortbl}
\definecolor{Gray}{rgb}{ 0.91, 0.91, 0.91 } 
\usepackage[flushleft]{threeparttable}
\usepackage{paralist}
\usepackage{tabularx}

\usepackage{eurosym} 

\usepackage{booktabs,caption}
\usepackage{graphicx}
\usepackage{tabularx}
\usepackage{threeparttablex}
\usepackage[normalem]{ulem}

\usepackage[a4paper,top=20mm, bottom=20mm, width=170mm]{geometry}

\usepackage{amsthm}
\usepackage{thmtools}
\usepackage{natbib}
\usepackage{subcaption}

\usepackage[labelformat=parens,labelsep=quad,skip=3pt]{caption}

\usepackage{lscape}
\usepackage{placeins}

\newtoggle{ShowQuestions}
\settoggle{ShowQuestions}{false}

\newcommand{\newterm}[1]{\textit{#1}}

\newtoggle{ShowQuestionsC}
\settoggle{ShowQuestionsC}{true}

\newtoggle{ShowQuestionsS}
\settoggle{ShowQuestionsS}{true}

\newtoggle{ShowToDo}
\settoggle{ShowToDo}{false}

\usepackage[none]{hyphenat} 

\usepackage[onehalfspacing]{setspace}
\usepackage{comment} 
\usepackage[colorlinks, linkcolor=red, citecolor=blue, urlcolor=blue]{hyperref}
\usepackage{bbold}

\usepackage{lipsum}  

\usepackage[flushmargin]{footmisc}

\usepackage[titletoc,title]{appendix}
\usepackage{titlesec}

\titleformat{\section}{\Large\bfseries}{\thesection .}{0.5em}{}
\titleformat{\subsection}{\large\bfseries}{\thesubsection }{0.5em}{}
\titleformat{\subsubsection}{\normalfont\bfseries}{\thesubsubsection}{0.5em}{}
\titlespacing{\section}{0pt}{\parskip}{\parskip}
\titlespacing{\subsection}{0pt}{0.25\baselineskip}{0.25\baselineskip}
\titlespacing{\subsubsection}{0pt}{0.25\baselineskip}{0.25\baselineskip}

\usepackage[utf8]{inputenc}
\usepackage[english]{babel}

\newtheorem{Proposition}{Proposition}

\usepackage{caption}
\usepackage{siunitx}
\usepackage[flushleft]{threeparttable}
\usepackage{amssymb}
\usepackage{titling}

\usepackage{xurl}

\newcommand{\SideBySideFigCaption}[5]{
      \begin{minipage}[c]{#4\textwidth}
        \includegraphics[width=\textwidth,keepaspectratio]{#1}
      \end{minipage}\hfill
      \begin{minipage}[c]{#5\textwidth}
        \caption{#2} \label{#3}
      \end{minipage}
}

\usepackage{hyperref}
\hypersetup{
    colorlinks=true,
    urlcolor=blue
}
{%
    \hspace*{-0.5cm} \begin{tabular}
}%
{%
    \end{tabular}\hspace*{0.5cm}
}

{%
    \hspace*{-0.5cm} \begin{threeparttable}
}%
{%
    \end{threeparttable}\hspace*{0.5cm}
}

\usepackage{placeins}
\renewcommand\footnotemark{}

\title{They were robbed! Scoring by the middlemost \\ to attenuate biased judging in boxing}
\author{\vspace{-2cm} Stuart Baumann \and Carl Singleton\thanks{\href{mailto:stuart@stuartbaumann.com}{stuart@stuartbaumann.com}; Corresponding author: \href{mailto:carl.singleton@stir.ac.uk}{carl.singleton@stir.ac.uk}, Economics Division, Stirling Management School, University of Stirling, Stirling, FK9 4LA, Scotland, UK. \newline We are grateful for comments and advice from Anwesha Mukherjee.\newline Declarations of interest: none}}
\date{ \today}

\begin{document}
\bibliographystyle{aea.bst}
\sloppy
\renewcommand\figurename{FIGURE}
\maketitle

\vspace{-1cm}\begin{abstract}

\noindent Boxing has a long-standing problem with biased judging, impacting both professional and Olympic bouts.
``Robberies'', where boxers are widely seen as being denied rightful victories, threaten to drive fans and athletes away from the sport.
To tackle this problem, we propose a minimalist adjustment in how boxing is scored: the winner would be decided by the majority of round-by-round victories according to the judges, rather than relying on the judges' overall bout scores.
This approach, rooted in social choice theory and utilising majority rule and middlemost aggregation functions, creates a coordination problem for partisan judges and attenuates their influence.
Our model analysis and simulations demonstrate the potential to significantly decrease the likelihood of a partisan judge swaying the result of a bout.
\end{abstract}
\vspace{0.2cm} 
\clearpage
\vspace{-0.5cm}\section{Introduction}
\sloppy 
\onehalfspacing
\noindent Boxing has a reputation for partisan and corrupt judging.
At the amateur level, some decisions in Olympic gold medal bouts have attracted criticism and ridicule, becoming boxing folklore, such as Roy Jones Jr.'s defeat in the 1988 (Seoul) light heavyweight final to a South Korean fighter \citep{Guardian2012}, and Joe Joyce's defeat in the 2016 (Rio de Janeiro) super heavyweight final \citep{Guardian2021,Telegraph2021}.
In professional boxing, there is longstanding suspicion about the integrity of judges (e.g., \citealp{USSenate2001}).
Recent perceived ``robberies'' include Haney Vs.\ Lomachenko \citep{Wainwright2023} and the first two editions of Alvarez Vs.\ Golovkin \citep{Reid2023}.

\noindent This short paper models the decisions of boxing judges and proposes an alternative scoring method that has the potential to significantly attenuate judge bias. 
Currently, scoring at the elite level is on a per-judge basis, with three judges usually employed for elite professional bouts and five at the Olympic amateur level. Judges score each round individually and then award their entire ``vote'' to the boxer who wins a majority of rounds.\footnote{This is a slight simplification as judges can award additional points for a given round based on knockdowns, fouls, or particularly dominant performances by one fighter.}
The bout is then awarded to the boxer receiving votes from a majority of judges. If neither boxer receives a majority, due to at least one tied scorecard among the judges, then the bout is a draw.
In this system, ``\newterm{aggregation over rounds and then judges}'', or \newterm{``majority judges rule''}, it is relatively straightforward for a judge to ensure their vote goes to their favoured boxer. They just need to award them half the rounds (i.e., 7 of 12 for a world championship level men's professional bout). They can do this while minimising backlash, by choosing the best rounds for their favoured boxer. 

\noindent The change we propose, ``\newterm{aggregation over judges and then over rounds}'', or ``\newterm{majority rounds rule}'', is for each round to be awarded based on the aggregate scores over all judges. Whoever wins the majority of rounds wins the bout, rather than whoever wins on a majority of the judges' scorecards. This represents a minimalist change to the scoring system in the sport, so that the aggregation of judges' scores is first between them within rounds, and then over rounds, rather than vice versa. The minor nature of this change is sufficient to introduce a significant coordination problem for a partisan judge, and may be acceptable among fans.\footnote{Fans tend to scrutinise, oppose and criticise even quite small changes to the rules of their beloved sports. A notable example from cricket is the LBW rule, which has been continually `improved' over the last century, often under opposition and criticism \citep{Kumar2022}.}

\noindent We focus on modelling the simplest practical case, with three judges, one of whom is biased in favour of one boxer. Under majority judges rule, a partisan judge can substantially increase the probability of a boxer winning despite being outnumbered by unbiased judges. Under majority rounds rule, even if the partisan judge awards a majority of rounds to a favoured boxer, then this will have no impact on the final outcome unless those rounds align with the decisions of the other judges. This coordination problem implies that, to achieve a high probability of victory for their favoured boxer, the partisan judge would have to award more rounds to their favoured boxer than in the current system. This exposes them to scrutiny and potential backlash, as boxing pundits and fans will often criticise poorly awarded rounds on judges' scorecards.\footnote{Criticism of judges in social and sports media is generally fiercest after bouts where a robbery is perceived to have happened. It often focuses on specific rounds where a judge's decision appears to be particularly poor. While authorities seldom intervene to order a rematch, judges may be stripped of their status and not employed again (e.g., \citealp{Mail2017}).} Our analysis and simulations of the model demonstrate that the scoring rule change could be highly effective in diminishing the incentives for biased judging in boxing and its influence.

\noindent Our proposed scoring rule is an application of majority rule and the middlemost aggregation function from social choice theory, which minimise the effective manipulability of outcomes by graders (e.g., \citealp{Arrow1963,Balinski2007,Young1974a,Young1974b}). This principle is already applied somewhat to the scoring in boxing, since the decision of the middlemost judge now determines the bout result. We suggest awarding bouts based on the aggregated middlemost round-by-round votes instead.

The prevalence of judging bias in combat sports has been documented in a growing literature of empirical academic papers (e.g., \citealp{LeeCorkAlgranati2002,HolmesMchaleZychaluk2024}). This behaviour was also described vividly in the recent judge-led independent investigation \citet{McLaren2022} report, which examined unethical conduct in Olympic boxing after being commissioned by the Association Internationale de Boxe Amateur (AIBA). While the report did propose improved appointment processes and training of judges, it did not explore how to make the incentives inherent in the judging process more resilient to biases and corruption.

Our goal of improving the incentives of judges in boxing is closely related to the focus of \citet{FrederiksenMachol1988}, who analysed the judging in sports like figure skating and dance, where judges need to decide between multiple competitors, a setting where Arrow's (\citeyear{Arrow1963}) theorem implies that all possible ways to combine judge preferences have some undesirable characteristics. \citeauthor{FrederiksenMachol1988} proposed a new method for aggregating judge scores for such situations that attenuates some of these issues. Their context though faced the problem of the Arrow Impossibility Theorem (social choice paradox), given there were more than two alternative outcomes in the contest. That theorem does not apply here for a boxing bout since it consists of just two competitors, only one winner, and potentially biased judges.

In general, this paper contributes to the vast operational research literature that either post hoc analyses changes to scoring rules and laws in sports or proposes new changes (for recent surveys see \citealp{Wright2014,Kendall2017}). Our work falls into the latter type of study, particularly where minimalist changes have been proposed that could still in theory substantially improve the fairness of sports outcomes. For instance, in the world's most popular sport, association football, recent contributions have used simulations to explore whether incentives and outcomes could be altered significantly under different tie-breaking rules in round-robin tournaments \citep{Csato2023,Csato2023b}, whether dynamic sequences in penalty shootouts could be fairer \citep{Csato2022}, and whether the allocation system for the additional slots of the expanded FIFA World Cup could be improved according to the stated goals of the organisers \citep{Krumer2023b}.
Finally this paper builds on a growing sports economics and management literature studying various incentive issues in boxing and other combat sports \citep{Akin2023,AmegashieKutsoati,Butler2023,Butler2023b,Duggan2002,Dietl2010,Tenorio2000}. However, to the best of our knowledge, the incentives of boxing judges have not yet been studied, given the scoring rules they face, despite a well-developed literature on the influences and implications of biased decision making by the referees and judges in other sports (e.g., \citealp{Dohmen2016,Bryson2021,Reade2022}), including other combat contests \citep{Brunello2023}).

The remainder of our short paper proceeds as follows. In Section 2, we setup a styled model of potentially biased judging in a boxing contest. Section 3 describes our analysis and discussion of the model. The detailed proofs of the main propositions regarding the scoring rules are presented in the Online Appendix, as are variations on the main results from simulating the model.

\section{The Model}

\noindent Consider a contest between two boxers of equal ability, in the \textit{B}lue and \textit{R}ed corners.
We assume each sequential round $t \in \{1,2,\dots,N\}$ of the contest has a true result, $\tau_t \in \{ B , R \}$, which is a binomial random variable with equal probability.

\noindent Each judge, $j\in\{1,2,3\}$, gets an i.i.d.\ signal, $x_{t,j} \in \{ B , R \}$, about the result of a round.
With probability $ \alpha \in (0,\frac{1}{2})$ this signal is the incorrect result, $x_{t,j}\neq \tau_t $,  while with probability $1-\alpha$ it is correct, $x_{t,j}\equiv \tau_t $.

Judges have a utility of:
\begin{align}
U &= S \mathbf{1}_\text{Blue wins} + G \mathbf{1}_\text{Red wins} - L \label{JudgeUtil} \ , & L &= \frac{\sum_{t=1}^{N} \mathbf{1}_{s_{t,j} \not= \tau_t}}{N} \ ,
\end{align}
where $S \geq 0$ and $G \geq 0$  represents the a judge's value from \textit{B}lue or \textit{R}ed winning respectively. $L$ refers to a backlash cost. 

We consider the case of two fair judges have $S = G = 0$. As these judge's utility does not depend on the bout's winner their optimal behaviour is to minimise backlash by awarding \textbf{fairly}, defined by choosing a round score of $s_{t,j}=B \iff x_{t,j} \equiv B$.\footnote{We analyse the case where all three judges are fair in Online~Appendix~\ref{NoJudgesBiased}.}
The third judge is \textbf{partisan} in favour of \textit{B}lue and so has $S > 0$ and $G = 0$.


\noindent Under majority judges rule, the middlemost judge scorecard determines the bout. Under majority rounds rule, the middlemost judge determines each round's winner, and then the middlemost round determines the bout. Judges award rounds separately and simultaneously.

\section{Analysis, Results, and Discussion}

\noindent The partisan judge ($j=1$) can minimise backlash by awarding rounds fairly.\footnote{If $S \equiv 0$, then this judge's actions will be congruent to the fair judges.}
Under majority judges rule, they can maximally increase the chance of \textit{B}lue winning the bout, while minimising backlash, by awarding $s_{t,1}=B$ in more than $\frac{N}{2}$ rounds.
Under majority rounds rule, their problem is more complex; a judge could award more than $\frac{N}{2}$ rounds to a boxer who then does not win them because the other judges disagreed.

\noindent If $S$ is low, however, then the expected backlash can be sufficient for the partisan judge to award rounds fairly. We can characterise the critical $\hat{S}$ where the partisan judge is indifferent between awarding fairly or gifting an additional round to \textit{B}lue. We find that this critical value is higher under majority rounds than majority judges rule, indicating that the former is more resilient to judge bias.

\renewcommand*{\proofname}{Sketch of Proof:}
\begin{Proposition}\label{MainPropOnThreeRoundBouts}
For three-round bouts, in which \textit{R}ed won a majority of rounds according to the true realisations, $\boldsymbol{\tau} = [\tau_1,\tau_2,\tau_3 ]$, the critical $\hat{S}$ is higher under majority rounds than majority judges rule, $\forall \alpha \in (0,\frac{1}{2})$.
\begin{proof}
We can calculate the probability of each fair judge awarding a round for \textit{B}lue, denoted by $q$, conditional on the signal seen by the partisan judge:
\begin{align}
q \vert (x_{t,1} \equiv B) &= (1 - \alpha)^2 + \alpha^2 \ , &  q \vert (x_{t,1} \equiv R) &= 2\alpha(1-\alpha) \ . \label{drawQ}
\end{align}
Under majority rounds rule, the number of fair judges awarding for \textit{B}lue in a particular round can be represented as drawing from a binomial distribution with probabilities as in Equation \ref{drawQ}. The survival function of this binomial, in conjunction with the decision of the partisan judge, is sufficient to infer the probability of \textit{B}lue winning the round. From the probabilities for each round, we can derive the optimal number of rounds for the partisan judge to award for \textit{B}lue.
Under majority judges rule, we can infer the probability of another scorecard being in favour of \textit{B}lue by combining the probabilities in Equation~\ref{drawQ} across rounds. We can use these probabilities to evaluate whether the partisan judge should award additional rounds such that \textit{B}lue wins on their card. Then we can derive, for each scoring rule, expressions for the critical $\hat{S}$ below which the partisan judge will award rounds fairly (see Online~Appendix~\ref{FullProof}). We find that there is a higher $\hat{S}$ under majority rounds rule, giving us the proposition.
\end{proof}
\end{Proposition}

\noindent Proposition~\ref{MainPropOnThreeRoundBouts} establishes that the majority rounds rule is more robust to partisan judging than the majority judges rule for three-round bouts. We numerically solve the model to establish the robustness of this result in longer bouts.\footnote{This is done along the lines discussed in the sketch proof for Proposition \ref{MainPropOnThreeRoundBouts}. The code for calculating partisan judge best responses and bout simulations are included in the online supplementary material.} We use a benchmark parametrisation of $\alpha = 0.1$, $S = 0.8$, three judges (one of whom is partisan), and $N=12$ rounds.\footnote{Importantly, Online~Appendix~\ref{NoJudgesBiased} shows that when all three judges are fair, the majority rounds rule is still more accurate than the majority judges rule in generating a deserving winner of the bout. Intuitively this occurs as there are more combinations of rounds that could be flipped to change the result. Consider a three-round bout with three fair judges and a $\mathbf{\tau}$ realisation of $\mathbf{[B,B,R]}$. Consider that two mistakes happened in judging the bout (in that for two round-judges the $x_{t,j}$ realisation differs from $\tau$). There are six possible pairs of $x_{t,j}$ values that can be flipped to change the result of the bout with the majority rounds rule. For one of the rounds where \textit{B}lue won, we need to flip two $x_{t,j}$ values and there are six combinations that achieve this. But there are twelve possible pairs of $x_{t,j}$ values that can be flipped to change the result of the bout with the majority judges rule. We need to flip two of the $x_{t,j}$ values awarded to \textit{B}lue on two different scorecards and there are twelve pairs of values that achieve this.}

\begin{figure}[ht!]
	\centering
 	\caption{Simulated Probability of \textit{B}lue winning, when both boxers truly won 6 of the 12 rounds, and 1 of the 3 judges favours \textit{B}lue}
		\includegraphics[width=1.0\textwidth]{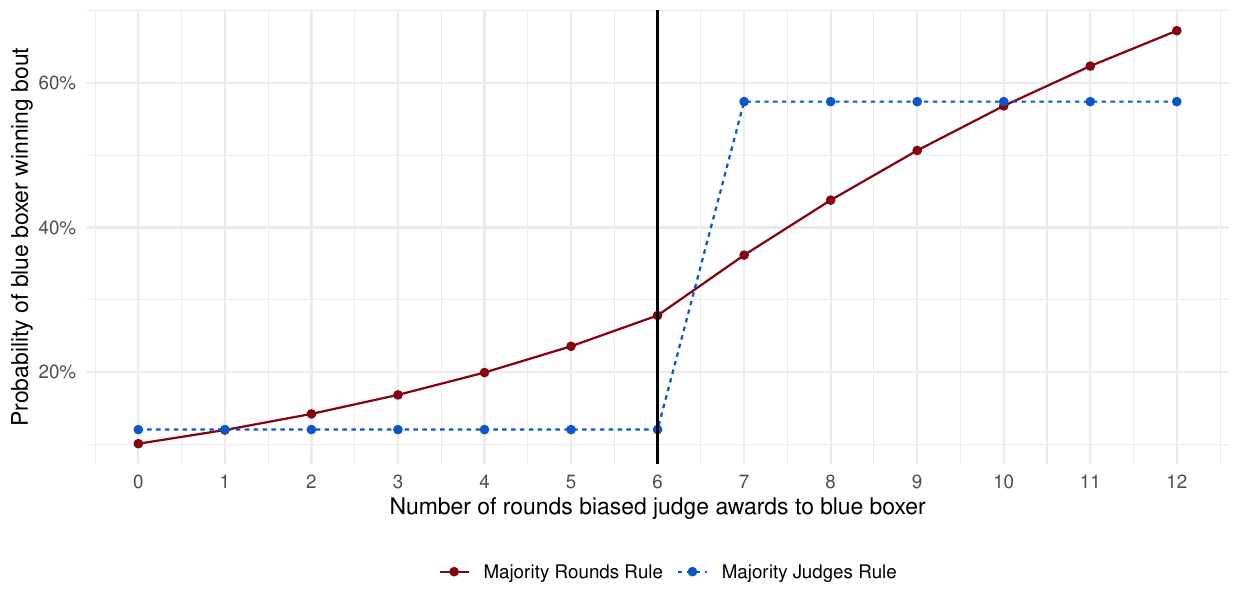}
	\label{fig:curve6}
\end{figure}

\noindent To demonstrate a partisan judge's decision making, Figure~\ref{fig:curve6} shows the probability of \textit{B}lue winning the bout, given they truly won 6 rounds, for each number of rounds the partisan judge awards them. Under majority judges rule, there is a sharp increase in the probability of \textit{B}lue winning if the partisan judge awards them more than 6 rounds.
If \textit{B}lue truly deserved to win 4 or 5 rounds, then, to award \textit{B}lue the win, the partisan judge only needs to risk the backlash associated with giving them 3 or 2 more rounds  on their scorecard.
In contrast, Figure~\ref{fig:curve6} shows that under majority rounds rule, a judge cannot secure a sharp increase in the probability of \textit{B}lue winning by giving them a small number of extra rounds; more rounds only gradually increase \textit{B}lue's chances.

\begin{figure}[ht!]
	\centering
 	\caption{Probability of a ``correct'' result depending on the number of rounds truly won by \textit{B}lue and how judges' scores are aggregated}
		\includegraphics[width=1.0\textwidth]{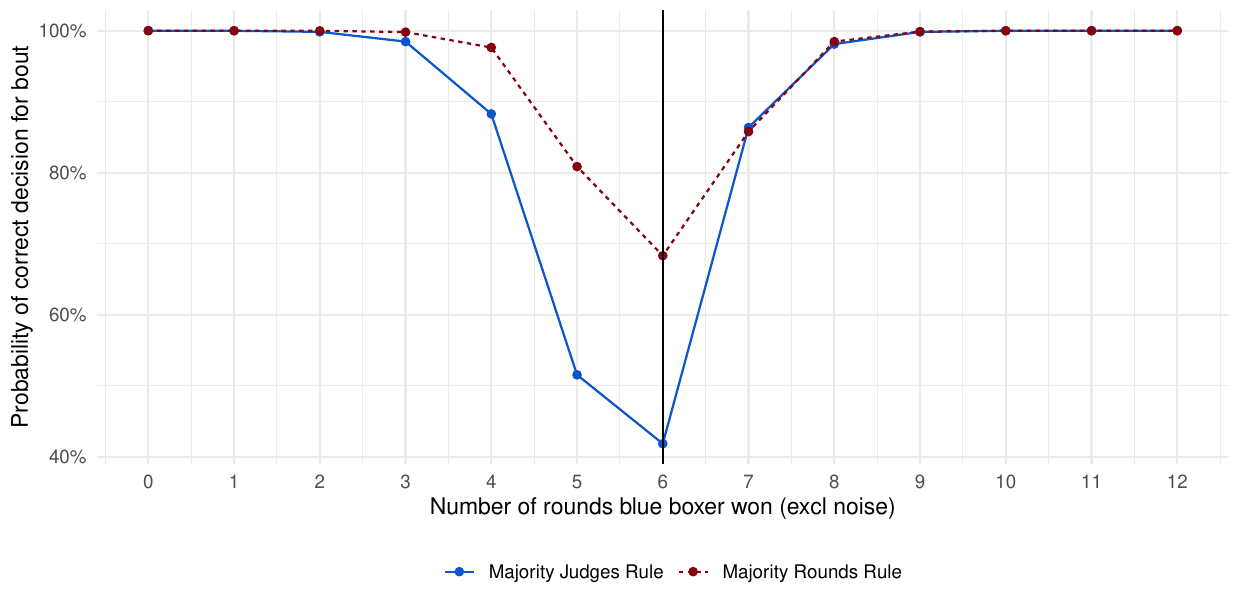}
	\label{fig:outcomes_of_bouts}
\end{figure}

\noindent Figure~\ref{fig:outcomes_of_bouts} shows the impact of these differing incentives for the partisan judge, from running a series of simulations and counting the proportion of times each boxer wins under the two scoring systems, conditional on the true number of rounds won by \textit{B}lue. When deciding the contest by majority judges, there is a high probability of erroneous results when \textit{B}lue truly won only 4-6 rounds.
When \textit{B}lue truly wins the most rounds, the partisan judge unduly helps to lock in a deserved victory, so there is not a large difference in the number of incorrectly awarded bouts.

Finally, in the majority rounds case, it can be noted from Figure~\ref{fig:outcomes_of_bouts} that the probability of the \textit{B}lue boxer winning always increases when the biased judge awards them more rounds. This is in contrast to the majority judges case, when a judge ceases to impact the result at the point at which they award a majority of their card to a boxer. For instance, consider a bout where the fair judge sees 10 rounds with $x_{t,1} \equiv \textit{B}$ and only two rounds are seen to be won by red. In this case, the biased judge may award additional rounds to \textit{B}lue to lock in a \textit{B}lue victory, while they would have no such incentive under majority judges rule.

This effect, however, does not tend to lead to a greater probability of an erroneous result under the majority rounds rule. The main reason for this is that the effect occurs in a context where \textit{B}lue has likely won a large majority of rounds and is likely to win the bout. The more important case is when a bout is more even and there is a sharp increase, under the majority judges rule, in the winning probability at the 7 round level in Figure~\ref{fig:outcomes_of_bouts}.

\begin{figure}[ht!]
	\centering
 \caption{Probability of each outcome depending on the number of rounds truly won by \textit{B}lue and how judges' scores are aggregated}
		\includegraphics[width=1.0\textwidth]{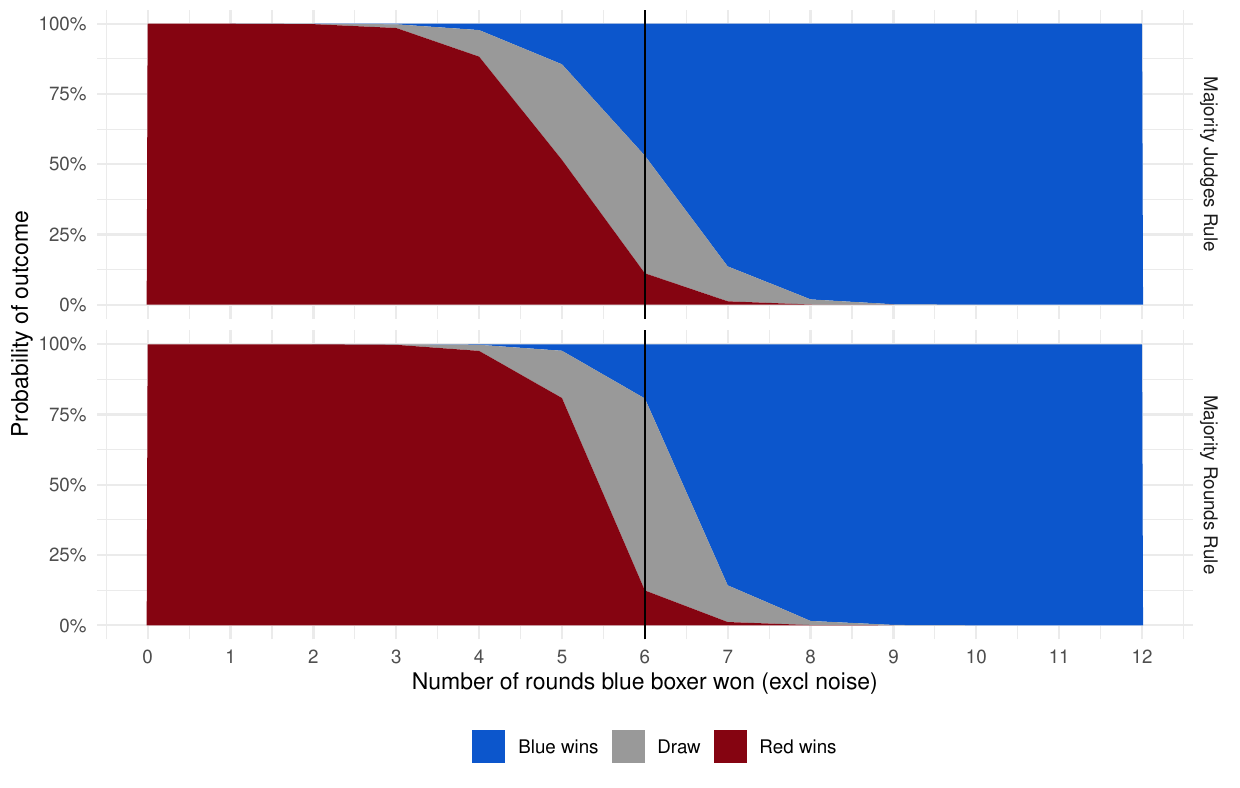}
	\label{fig:outcomes_of_bouts2}
\end{figure}

This point can be seen in Figure~\ref{fig:outcomes_of_bouts2}, which shows the probability of each possible outcome on the y-axis and the number of rounds \textit{B}lue truly won (excluding noise) on the x-axis.
Under majority judges rule (top panel), in evenly matched bouts, where the true result is a draw, \textit{B}lue wins 47.0\% and \textit{R}ed wins 11.2\%.
When evenly matched bouts are awarded under majority rounds rule (bottom panel), \textit{B}lue wins 13.4\% and \textit{R}ed wins 19.3\%.

Figure~\ref{fig:outcomes_of_bouts2} also shows the frequencies where one boxer wins despite the other deserving outright victory, e.g., the blue area to the left of the vertical black line.
Under majority judges rule, it is more likely for an erroneous victory to be in favour of \textit{B}lue than \textit{R}ed; in this parametrisation, a robbery in favour of \textit{B}lue is 12.5 times more likely than a robbery in favour of \textit{R}ed.
Under majority rounds rule, the likelihood of a robbery is still in \textit{B}lue's favour, by a multiple of 1.99, because there is still some incentive for the partisan judge to favour \textit{B}lue.
But this scoring system can substantially attenuate \textit{B}lue's advantage from the presence of a partisan judge.
There are also fewer robberies in absolute terms.

\noindent For robustness, the Online Appendices demonstrate extensions and checks on our analysis.
Appendix~\ref{OtherParameterisations} considers simulations with alternative parametrisations of the benchmark model, and, in Appendices \ref{RepeatEverythingForWomensPro}-\ref{RepeatEverythingForWomensAmateur} we repeat the analysis for setups consistent with women's professional, men's Olympic, and women's Olympic boxing, respectively (i.e., different numbers of rounds and judges). 
The results of all these extensions support our key findings: deciding bouts by majority rounds, compared with by majority judges, makes it less likely that a partisan judge sways the outcome of a bout.

\FloatBarrier


\clearpage
\section*{Declarations}
The authors did not receive financial support from any organisation for the submitted work. The authors have no competing interests to declare that are relevant to the content of this article.
\clearpage
\setlength{\parskip}{.5\baselineskip} 
\begin{raggedright}
\setcounter{secnumdepth}{-2}
\bibliography{short_version2e}
\end{raggedright}

\clearpage

\begin{center}
\renewcommand*{\thefootnote}{\fnsymbol{footnote}}
\setcounter{footnote}{1}
\normalsize 
\Large{They were robbed! Scoring by the middlemost \\ to attenuate biased judging in boxing}
\vspace{0.2cm}

\textbf{\Large{Online Appendix}}\\
\vspace{0.2cm}


\end{center}

\singlespacing

\begin{appendices}
\setcounter{secnumdepth}{2}
\titleformat{\section}{\large\bfseries}{\appendixname~\thesection .}{0.5em}{}
\renewcommand\thetable{\thesection\arabic{table}}
\renewcommand\thefigure{\thesection\arabic{figure}}
\renewcommand\theequation{\thesection\arabic{equation}}

\FloatBarrier

\section{Full proof of Proposition~\ref{MainPropOnThreeRoundBouts}}\label{FullProof}
\renewcommand\thetable{\thesection\arabic{table}}
\renewcommand\thefigure{\thesection\arabic{figure}}
\renewcommand{\theequation}{\Alph{section}.\arabic{equation}}
\setcounter{equation}{0}
\setcounter{table}{0}
\setcounter{figure}{0}


\noindent To simplify, we assume there are three rounds. Consequently, there are four information sets that the partisan judge can receive before they choose how many rounds to award to \textit{B}lue. These are \{ BBB, BBR, BRR, RRR \}, which are the signals of each round of the bout after sorting round results.\footnote{Note, in both scoring systems, no distinctions are made as to when in the bout a particular round result occurred. Therefore, a bout with true result BRB is the same as a bout with true result BBR, and we reorder round results to simplify the analysis.}

\subsection{Majority Judges Rule}

\noindent We will first analyse the majority judges case. We will denote that the number of rounds that \textit{B}lue truly won is $b_T$. Conditioning on $b_T$, the number of rounds that the \textit{B}lue boxer wins on the fair judge's scorecare is $x + y$, where $x \sim \text{Binomial}(b_T, 1-\alpha)$ and $y \sim \text{Binomial}(N - b_T, \alpha)$. The probability that the \textit{B}lue boxer wins on a fair judge's card is, therefore:
\begin{align}
\text{Prob}( \text{Blue wins} \vert b_T) &= \sum_{x=0}^3 f_x(x) G_y(2 - x) \label{MajJudgesAppFunc1}
\end{align}
where $f_x(\cdot)$ is the probability mass function of $\text{Binomial}(b_T, 1-\alpha)$ and $G_y(\cdot)$ is the survival function of $\text{Binomial}(N - b_T, \alpha)$. 

\noindent The partisan judge cannot condition on $b_T$, however, as they cannot observe it. They can assign probabilities to the various possibilities of $b_T$ after conditioning on their observation of $b$. We can work out $\text{Prob}(b \vert  b_T)$ using the following expression that is similar to Equation \eqref{MajJudgesAppFunc1}:
\begin{align}
\text{Prob}( \text{Blue wins $b$ rounds} \vert b_T) &= \sum_{x=0}^3 f_x(x) g_y(b - x) \label{MajJudgesAppFunc2}
\end{align}
where $g_y(\cdot)$ is the probability mass function of $\text{Binomial}(N - b_T, \alpha)$. We can use this expression in conjunction with Bayes rule to derive an expression for  $\text{Prob}(b_T \vert  b)$.

\noindent We use these to calculate the probability of another judge's scorecard being in favour of \textit{B}lue for each information set that the partisan judge observes. We use $C$ to denote this probability:
\begin{align}
C \vert b &= \sum_{b_T=0}^3 \text{Prob}(b_T \vert b) \times \text{Prob}( \text{Blue wins} \vert b_T)\label{MajJudgesAppFunc3}
\end{align}

\noindent The utilities available at each information set (where the first column shows the information set for each row) and action (the second to fifth columns) are shown in Table~\ref{UtilitiesFOrEachActionAndInfoSet}.

\begin{table}[h!]
\begin{adjustbox}{width=\columnwidth,center}
\begin{tabular}{|l|l|l|l|l|}
\hline
    & Award BBB & Award BBR & Award BRR & Award RRR \\ \hline
BBB & $ S(1-(1-C \vert BBB)^2)  - \alpha $
    & $ S(1-(1-C \vert BBB)^2)  - \frac{2\alpha + (1-\alpha)}{3}$
    & $S(C \vert BBB)^2  - \frac{\alpha + 2(1-\alpha)}{3}$        
    & $S(C \vert BBB)^2  - (1-\alpha)$        \\
BBR & $ S(1-(1-C \vert BBR)^2)  - \frac{2\alpha + (1-\alpha)}{3}$        
    & $ S(1-(1-C \vert BBR)^2)  - \alpha $         
    & $S(C \vert BBR)^2  - \frac{2\alpha + (1-\alpha)}{3}$        
    & $S(C \vert BBR)^2  - \frac{\alpha + 2(1-\alpha)}{3}$        \\
BRR & $ S(1-(1-C \vert BRR)^2)  - \frac{\alpha + 2(1-\alpha)}{3}$        
    & $ S(1-(1-C \vert BRR)^2)  - \frac{2\alpha + (1-\alpha)}{3}$         
    & $S(C \vert BRR)^2  - \alpha $        
    & $S(C \vert BRR)^2  - \frac{2\alpha + (1-\alpha)}{3}$        \\
RRR & $ S(1-(1-C \vert RRR)^2)  - (1-\alpha)$        
    & $ S(1-(1-C \vert RRR)^2)  - \frac{\alpha + 2(1-\alpha)}{3}$         
    & $S(C \vert RRR)^2  - \frac{2\alpha + (1-\alpha)}{3}$        
    & $S(C \vert RRR)^2  - \alpha $        \\
\hline
\end{tabular}
\end{adjustbox}
\caption{Utilities for each action and information set in the basic model}\label{UtilitiesFOrEachActionAndInfoSet}
\end{table}

\noindent The best responses to seeing BBB and BBR are to award BBB and BBR, respectively. This minimises backlash while still awarding the card of the partisan judge to the favoured \textit{B}lue.
If the partisan judge sees BRR, however, then they need to choose between BBR (which awards the card to \textit{B}lue) and BRR (which minimises expected backlash). The condition for awarding rounds BRR as preferred to awarding rounds BBR is:
\begin{align}
\widehat{S}_\text{Maj. Jgs, BRR} &\leq \frac{1 - 2\alpha}{6(C \vert BRR)(1 - (C \vert BRR))}\notag\\
 &\leq (1-2\alpha) \div \Bigg[192\alpha \left(1- \alpha\right) \left(32 \alpha \left(\alpha - 1\right) \left(0.5 \alpha^{4} - 1 \alpha^{3} + 0.875 \alpha^{2} - 0.375 \alpha + 0.125\right) + 1\right)\notag\\
 &\hspace{3cm} \times \left(0.5 \alpha^{4} - 1 \alpha^{3} + 0.875 \alpha^{2} - 0.375 \alpha + 0.125\right)\Bigg]
\label{marginalBacklashBRRmajorityJudges}
\end{align}
For a simple example, when $\alpha = 0.1$, this expression becomes approximately $1.09$.

\noindent If the partisan judge sees RRR, then they need to choose between BBR (which awards the card to \textit{B}lue) and RRR (which minimises backlash). The condition for RRR to be perferred to BBR is:
\begin{align}
\widehat{S}_\text{Maj. Jgs, RRR} &\leq \frac{1 - 2\alpha}{3(C \vert RRR)(1 - (C \vert RRR))}\notag\\
 &\leq (1 - 2\alpha) \div \Bigg[ 48 \alpha^{2} \left(1 - \alpha\right)^{2} \left(1 - 16 \alpha^{2} \left(1 - \alpha\right)^{2} \cdot \left(\alpha^{2} - \alpha + 0.75\right)\right) \left(\alpha^{2} - \alpha + 0.75\right) \Bigg]
 \label{marginalBacklashRRRmajorityJudges}
\end{align}
For a simple example, when $\alpha = 0.1$, the right-hand-side here becomes approximately $12.14$.

It makes intuitive sense that the critical level of \textit{B}lue winning utility will be higher here than in Equation~\eqref{marginalBacklashBRRmajorityJudges}, as they need to award two more rounds than they believe \textit{B}lue won (so higher backlash) and there is lower odds of at least one other judge awarding in favour of \textit{B}lue (so less chance that partisan judging will deliver a victory).

\subsection{Majority Rounds Rule}

\noindent We can derive the following probabilities of \textit{B}lue winning a round conditional on what the partisan judge observes and does:\footnote{For instance, we can work out $q_1$ and $q_3$ as follows. If we see B, then there is $(1-\alpha)$ chance that B is the true state and $\alpha$ chance that $R$ is the true state. If B is the true state, then there is a chance $\alpha$ that a fair judge sees R and a $1-\alpha$ chance they see B. If R is the true state, then there is a chance $\alpha$ that a fair judge sees B and a $1-\alpha$ chance they see R. The chance that B wins on at least one other scorecard (and thus wins the round) is therefore $(1-\alpha)\left[\underbrace{1 - \alpha^2}_\text{$B$ true state} \right] + \alpha \left[ \underbrace{1 - (1-\alpha)^2}_\text{$R$ true state} \right] = 1-\alpha(1-\alpha)$. We can work out $q_2$, $q_3$ and $q_4$ analogously.}
\begin{align}
\text{Partisan judge observes B and does B:} && q_1 &= 1-\alpha(1-\alpha)\\
\text{Partisan judge observes R and does B:} && q_2 &= 3\alpha - 3\alpha^2\\
\text{Partisan judge observes B and does R:} && q_3 &= 1 - 3\alpha + 3\alpha^2\\
\text{Partisan judge observes R and does R:} && q_4 &= \alpha(1-\alpha)
\end{align}

\noindent At this point, we define the function $f$ to denote the probability of getting at least two realisations from 3 binomial distribution trials with probabilities $p_1,p_2,p_3$:
\begin{align}
f(p_1,p_2,p_3) = p_1p_2(1-p_3) + p_1(1-p_2)p_3 + (1-p_1)p_2p_3 + p_1p_2p_3
\end{align}
Using this function, we can write the utilities available at each information set (where the first column shows the information set for each row) and action (the second to fifth columns), shown in Table~\ref{UtilitiesFOrEachActionAndInfoSet2}.
\begin{table}[h!]
\begin{adjustbox}{width=\columnwidth,center}
\begin{tabular}{|l|l|l|l|l|}
\hline
    & Award BBB & Award BBR & Award BRR & Award RRR \\ \hline
BBB & $ Sf(q_1,q_1,q_1) - \frac{3\alpha + 0(1-\alpha)}{3}$ 
    & $ Sf(q_1,q_1,q_3) - \frac{2\alpha + 1(1-\alpha)}{3}$
		& $ Sf(q_1,q_3,q_3) - \frac{1\alpha + 2(1-\alpha)}{3}$
		& $ Sf(q_3,q_3,q_3) - \frac{0\alpha + 3(1-\alpha)}{3}$	\\
BBR & $ Sf(q_1,q_1,q_2) - \frac{2\alpha + 1(1-\alpha)}{3}$ 
    & $ Sf(q_1,q_1,q_4) - \frac{3\alpha + 0(1-\alpha)}{3}$
		& $ Sf(q_1,q_3,q_4) - \frac{2\alpha + 1(1-\alpha)}{3}$
		& $ Sf(q_3,q_3,q_4) - \frac{1\alpha + 2(1-\alpha)}{3}$	\\
BRR & $ Sf(q_1,q_2,q_2) - \frac{1\alpha + 2(1-\alpha)}{3}$ 
    & $ Sf(q_1,q_2,q_4) - \frac{2\alpha + 1(1-\alpha)}{3}$
		& $ Sf(q_1,q_4,q_4) - \frac{3\alpha + 0(1-\alpha)}{3}$
		& $ Sf(q_3,q_4,q_4) - \frac{2\alpha + 1(1-\alpha)}{3}$	\\
RRR & $ Sf(q_2,q_2,q_2) - \frac{0\alpha + 3(1-\alpha)}{3}$ 
    & $ Sf(q_2,q_2,q_4) - \frac{1\alpha + 2(1-\alpha)}{3}$
		& $ Sf(q_2,q_4,q_4) - \frac{2\alpha + 1(1-\alpha)}{3}$
		& $ Sf(q_4,q_4,q_4) - \frac{3\alpha + 0(1-\alpha)}{3}$	\\
\hline
\end{tabular}
\end{adjustbox}
\caption{Utilities for each action and information set in the basic model under majority rounds rule }\label{UtilitiesFOrEachActionAndInfoSet2}
\end{table}

\noindent Similar to the case in Table~\ref{UtilitiesFOrEachActionAndInfoSet}, the utility on the diagonal is better than the utility from awarding more rounds than this to \textit{R}ed. This implies that if the partisan judge sees BBB, then they should award rounds as BBB.
We can solve for the levels of $S$ at which the partisan judge prefers to award rounds fairly rather than giving additional rounds to \textit{B}lue. Starting at the BBR information set:
\begin{align}
\widehat{S}_\text{Maj. Rds, BBR} &\leq 
\frac{1 - 2 \alpha}{12\alpha^2 \left[ \alpha^4 - 3\alpha^3 +4\alpha^2 -3\alpha + 1 \right]}
 \label{marginalBacklashBBRmajorityRounds}
\end{align}
For the BRR information set, the partisan judge could do BBR or BBB rather than the fair result of BRR. We derive the critical $S$ for both and can determine that a partisan judge will deviate to BBR at a higher $S$ value than they would deviate to BBB. Hence, we below report the threshold above which the partisan judge will not deviate to BBR:
\begin{align}
&\widehat{S}_\text{Maj. Rds, BRR} \leq \notag\\ 
&\frac{1 - 2 \alpha}{6\alpha \left[ -2\alpha^5 + 6\alpha^4 - 8\alpha^3 + 6\alpha^2 - 3\alpha + 1  \right]}
 \label{marginalBacklashBRRmajorityRounds}
\end{align}
Finally, for the RRR information set, the partisan judge could do BRR, BBR or BBB, rather than the fair result of RRR. We can establish that if $\alpha$ is near zero, then the partisan judge will deviate to BBB at a higher $S$ than they would deviate to the other options. When $\alpha$ is near (but below) $0.5$, then they will deviate to BBR at a higher $S$ than they would deviate to the other options. As a result, we have the critical $S$ value:
\begin{align}
&\widehat{S}_\text{Maj. Rds, RRR} \leq \min \Large[\notag\\
&\frac{1 - 2 \alpha}{4 \alpha^{2} \left(13\alpha^4 - 39\alpha^3 + 45\alpha^2 - 25\alpha + 6\right)},\notag\\
&\frac{1 - 2 \alpha}{ 6 \alpha^{2} \left(  4\alpha^4 - 12\alpha^3 + 15\alpha^2 - 10\alpha + 3 \right)}
 \Large]
 \label{marginalBacklashRRRmajorityRounds}
\end{align}

\begin{figure}
	\centering
		\includegraphics[width=1.00\textwidth]{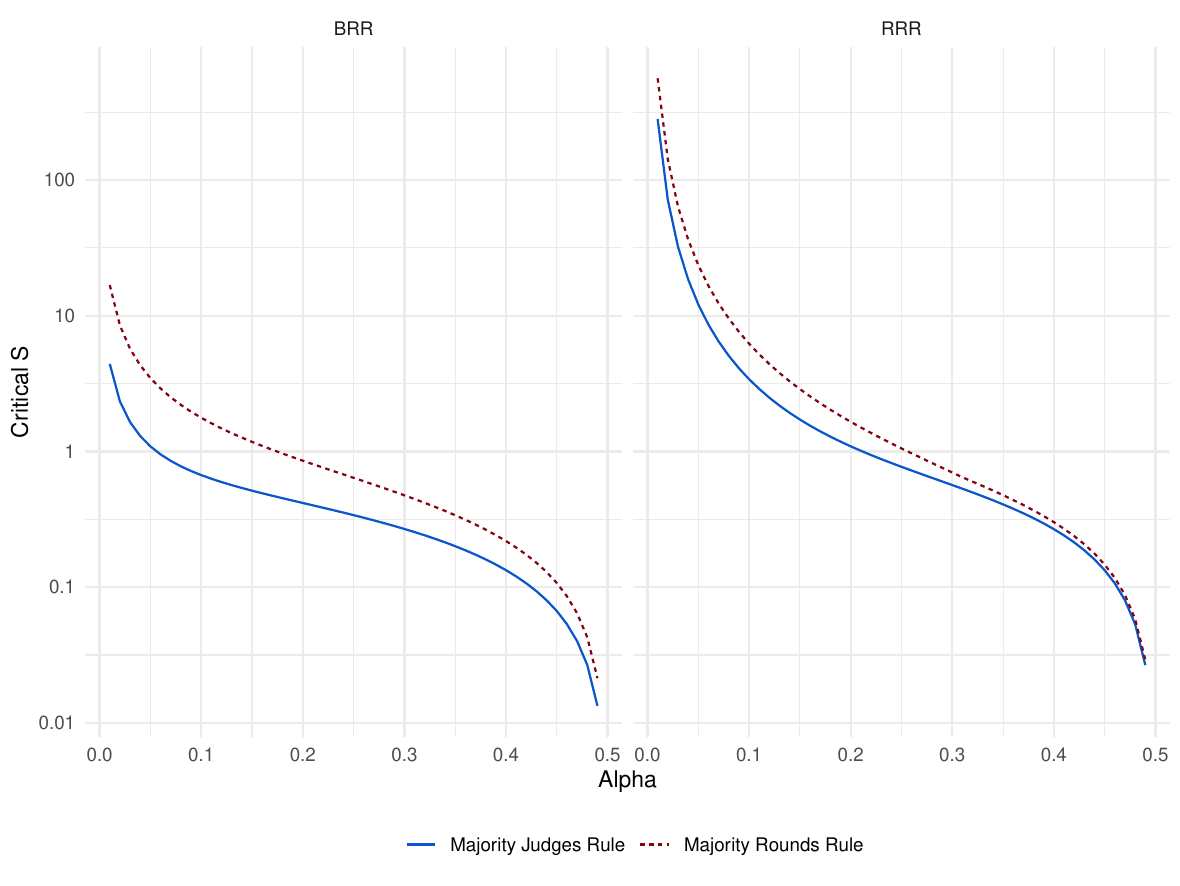}
	\caption{Critical $S$ below which partisan judges do award fairly for three-round bouts with three judges in the BRR and RRR information sets}
	\label{fig:Critical_BS}
\end{figure}

\noindent Now we can summarise Equations \eqref{marginalBacklashBRRmajorityJudges}, \eqref{marginalBacklashRRRmajorityJudges}, \eqref{marginalBacklashBBRmajorityRounds}, \eqref{marginalBacklashBRRmajorityRounds} and \eqref{marginalBacklashRRRmajorityRounds} with a chart of $\alpha$ against the critical $S$ ratio below which the partisan judge does not mis-award rounds. This is shown in Figure~\ref{fig:Critical_BS}.

\noindent In all cases where \textit{R}ed wins more rounds, the majority rounds rule has a higher $S$ value at which the partisan judge is indifferent to awarding fairly and giving more rounds to \textit{B}lue. This indicates that the majority rounds rule is more robust to partisan judging.
\FloatBarrier
\clearpage
\section{No judges are partisan}\label{NoJudgesBiased}
\renewcommand\thetable{\thesection\arabic{table}}
\renewcommand\thefigure{\thesection\arabic{figure}}
\renewcommand{\theequation}{\Alph{section}.\arabic{equation}}
\setcounter{equation}{0}
\setcounter{table}{0}
\setcounter{figure}{0}

\begin{figure}[ht!]
    \centering
	\caption{Probability of a ``correct'' result depending on the number of rounds truly won by \textit{B}lue and how judges' scores are aggregated: the case of no partisan judges}
		\includegraphics[width=0.85\textwidth]{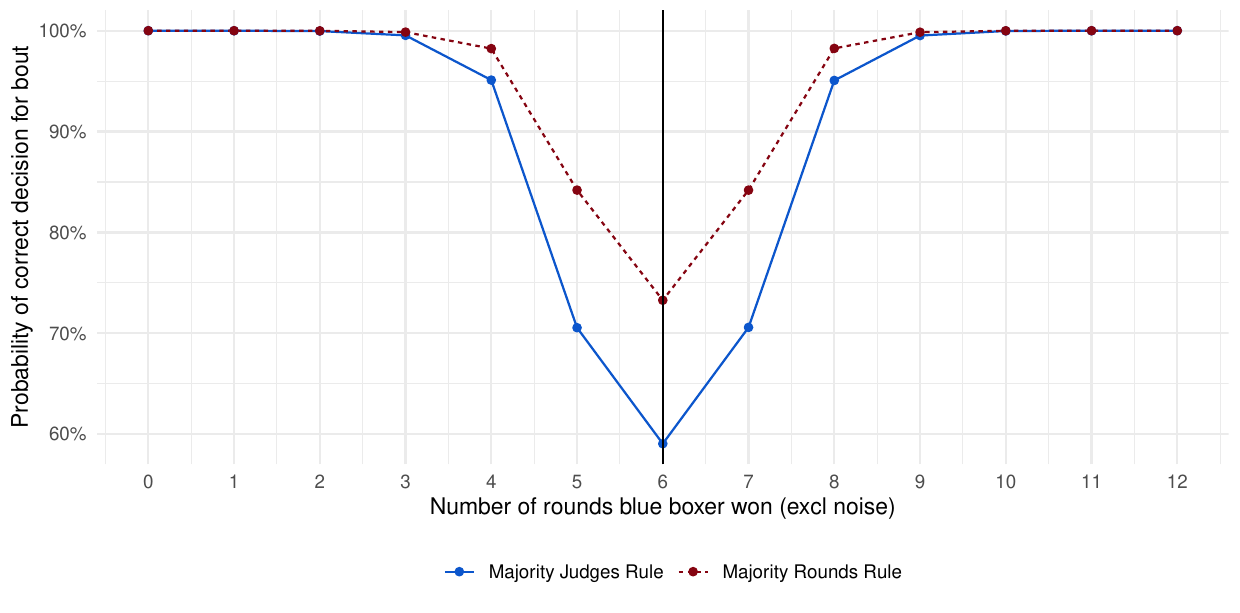}
	\label{fig:outcomes_of_boutsHonesty}
\end{figure}

\noindent When no judges are partisan, the proposed change, from awarding bouts by majority judges to majority rounds, reduces the probability of bouts being wrongly decided. This can be seen in Figure~\ref{fig:outcomes_of_boutsHonesty}, which is comparable to Figure~\ref{fig:outcomes_of_bouts}, with the same parametrisation, but reflects the case where all three judges are fair.

\FloatBarrier
\clearpage
\section{Other parametrisations}\label{OtherParameterisations}
\renewcommand\thetable{\thesection\arabic{table}}
\renewcommand\thefigure{\thesection\arabic{figure}}
\renewcommand{\theequation}{\Alph{section}.\arabic{equation}}
\setcounter{equation}{0}
\setcounter{table}{0}
\setcounter{figure}{0}

\noindent We consider different model parametrisations, to demonstrate the extent to which the qualitative results of this paper may vary.

\noindent \textbf{High disagreement between different judges}

\begin{figure}[ht!]
	\centering
 	\caption{Probability of favoured boxer winning in one bout where both boxers won 6 rounds (in the absence of noise)}
		\includegraphics[width=0.85\textwidth]{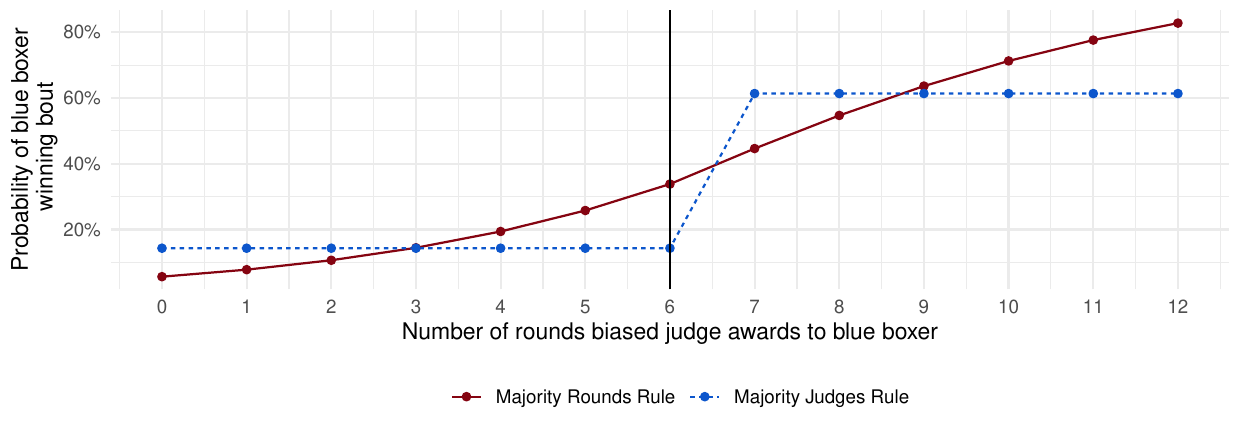}
	\label{fig:curve6_disagreement}
\end{figure}

\begin{figure}[ht!]
	\centering
 	\caption{Probability of correct result for bout}
		\includegraphics[width=0.85\textwidth]{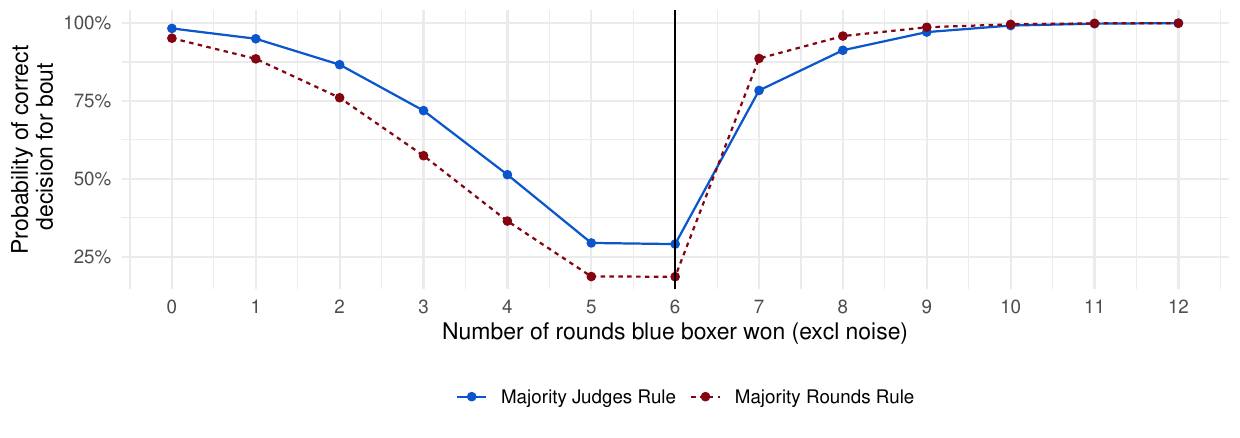}
	\label{fig:outcomes_of_bouts_disagreement}
\end{figure}

\begin{figure}[ht!]
	\centering
 	\caption{Probability of each outcome for bout}
		\includegraphics[width=0.8\textwidth]{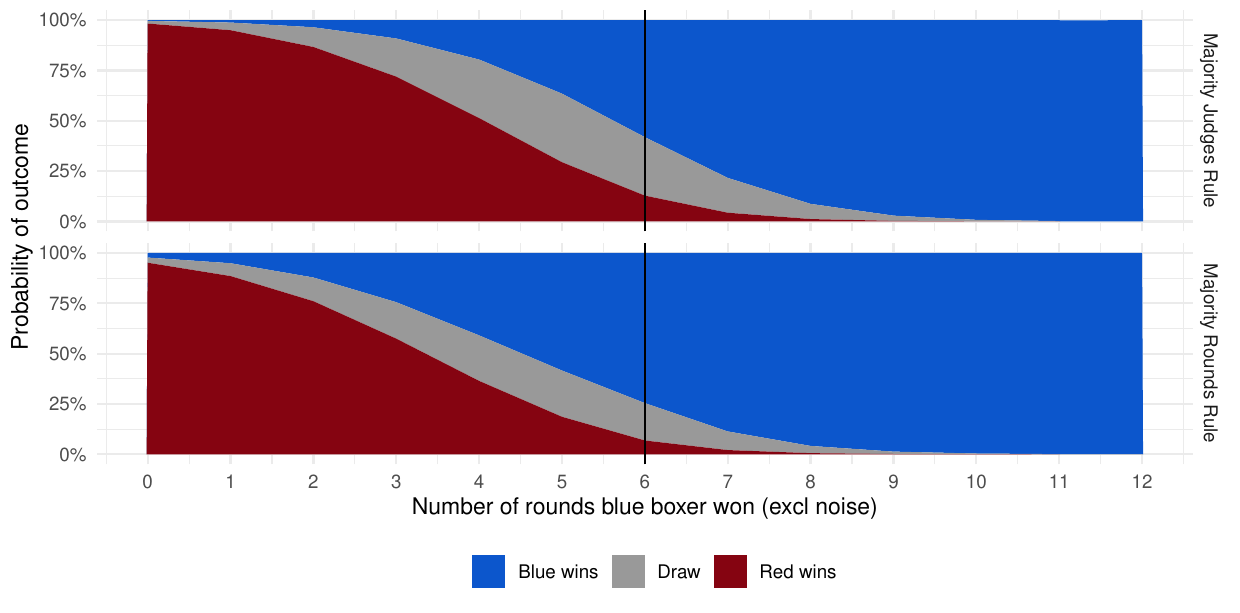}
	\label{fig:outcomes_of_bouts2_disagreement}
\end{figure}

\noindent We increase the noise variance, such that the different judges disagree more often about the outcome of a round. Specifically, we increase $\alpha = 0.2$ and leave the other parameters as they were in the main body of the paper. Figures \ref{fig:curve6}-\ref{fig:outcomes_of_bouts2} are reproduced below for this new parametrisation as Figures~\ref{fig:curve6_disagreement}- \ref{fig:outcomes_of_bouts2_disagreement}.
\vspace{-0.5cm}
\vspace{-0.5cm}

\clearpage

\noindent \textbf{High degree of favouritism}

\begin{figure}[ht!]
	\centering
 	\caption{Probability of favoured boxer winning in one bout where both boxers won 6 rounds (in the absence of noise)}
		\includegraphics[width=0.85\textwidth]{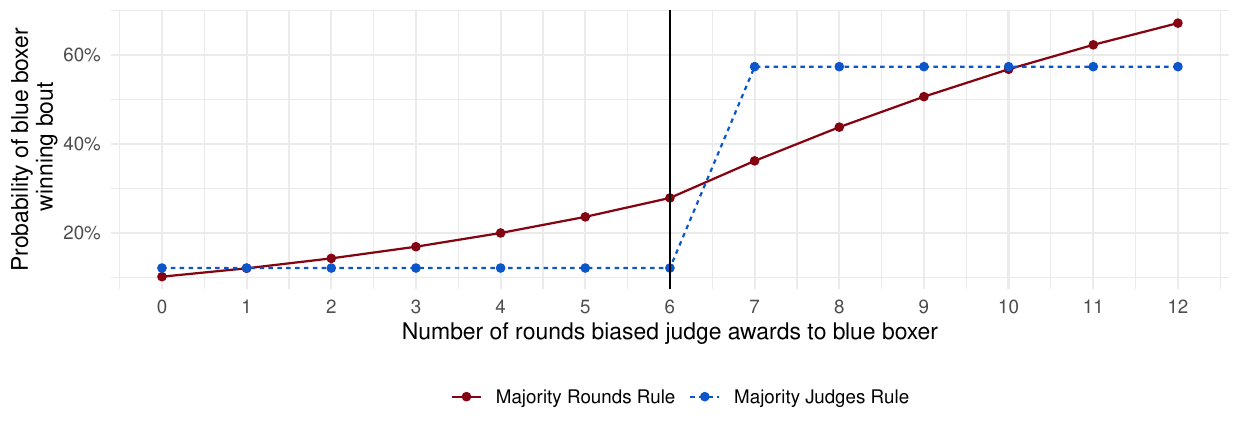}
	\label{fig:curve6_fav}
 \end{figure}

\begin{figure}[ht!]
	\centering
 	\caption{Probability of correct result for bout}
		\includegraphics[width=0.85\textwidth]{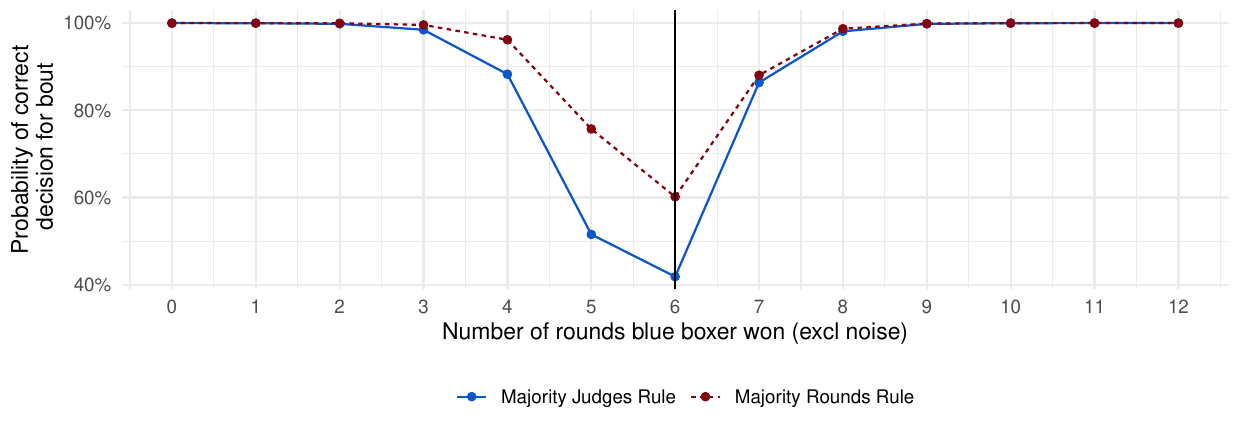}

	\label{fig:outcomes_of_bouts_fav}
 \end{figure}

\begin{figure}[ht!]
	\centering
 	\caption{Probability of each outcome for bout}
		\includegraphics[width=0.85\textwidth]{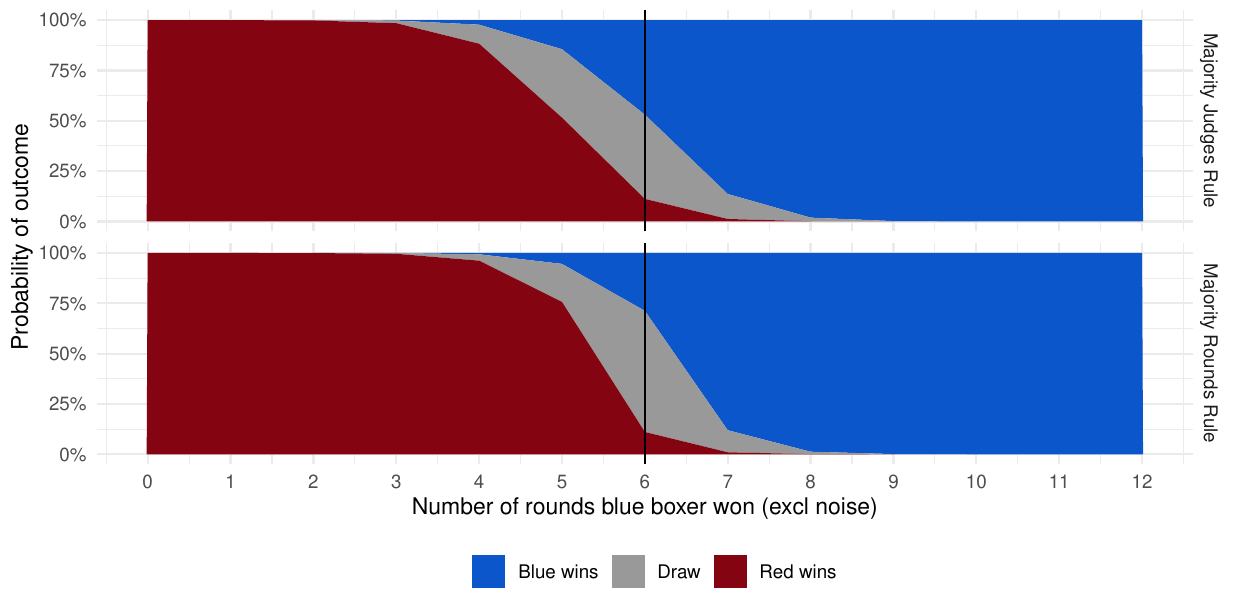}
	\label{fig:outcomes_of_bouts2_fav}
\end{figure}

\noindent We increase $S$ to 1.0 and leave other parameter values as they are in the main body of the paper. Figures \ref{fig:curve6}-\ref{fig:outcomes_of_bouts2} are reproduced below for this new parametrisation as Figures~\ref{fig:curve6_fav}-\ref{fig:outcomes_of_bouts2_fav}

 \vspace{-0.5cm}
 \vspace{-0.5cm}

\clearpage

\section{Women's professional boxing}\label{RepeatEverythingForWomensPro}
\renewcommand\thetable{\thesection\arabic{table}}
\renewcommand\thefigure{\thesection\arabic{figure}}
\renewcommand{\theequation}{\Alph{section}.\arabic{equation}}
\setcounter{equation}{0}
\setcounter{table}{0}
\setcounter{figure}{0}

\begin{figure}[ht!]
	\centering
 	\caption{Probability of favoured boxer winning in one bout where both boxers won 6 rounds (in the absence of noise) - Women's professional boxing}
		\includegraphics[width=0.85\textwidth]{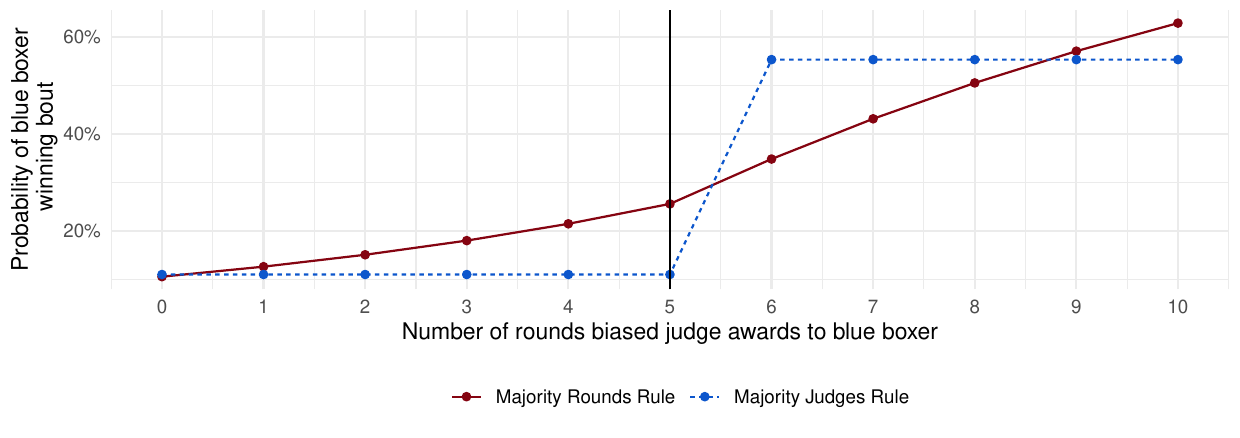}
	\label{fig:curve6_womenpro}
\end{figure}

\begin{figure}[ht!]
	\centering
 	\caption{Probability of correct result for bout - Women's professional boxing}
		\includegraphics[width=0.85\textwidth]{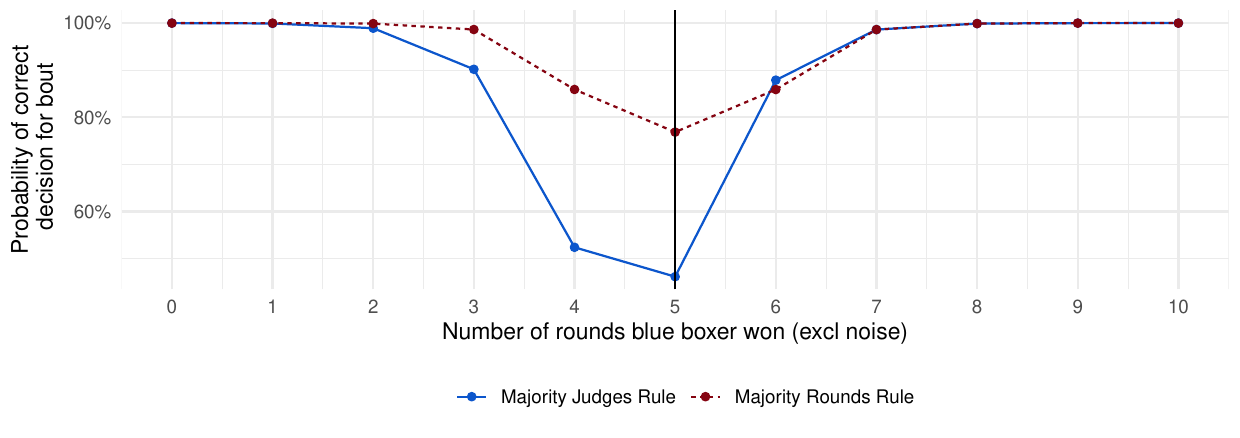}
	\label{fig:outcomes_of_bouts_womenpro}
\end{figure}

\begin{figure}[ht!]
	\centering
 	\caption{Probability of each outcome for bout - Women's professional boxing}
		\includegraphics[width=0.85\textwidth]{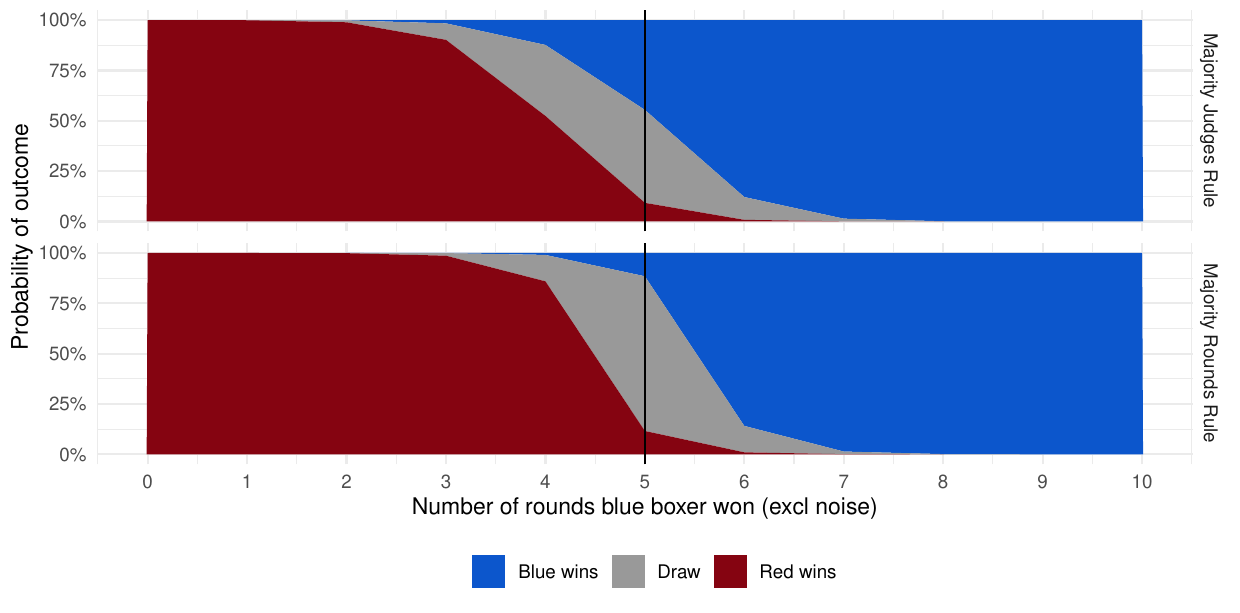}
	\label{fig:outcomes_of_bouts2_womenpro}
\end{figure}

\noindent In women's professional boxing, there are generally 10 rounds and 3 judges. Figures~\ref{fig:curve6}- \ref{fig:outcomes_of_bouts2} are reproduced below for women's professional boxing as Figures~\ref{fig:curve6_womenpro}-\ref{fig:outcomes_of_bouts2_womenpro}, with otherwise identical parametrisations.

 \vspace{-0.5cm}
 \vspace{-0.5cm}

\clearpage

\section{Men's Olympic amateur boxing}\label{RepeatEverythingForMensAmateur}
\renewcommand\thetable{\thesection\arabic{table}}
\renewcommand\thefigure{\thesection\arabic{figure}}
\renewcommand{\theequation}{\Alph{section}.\arabic{equation}}
\setcounter{equation}{0}
\setcounter{table}{0}
\setcounter{figure}{0}

\begin{figure}[ht!]
	\centering
		\includegraphics[width=0.85\textwidth]{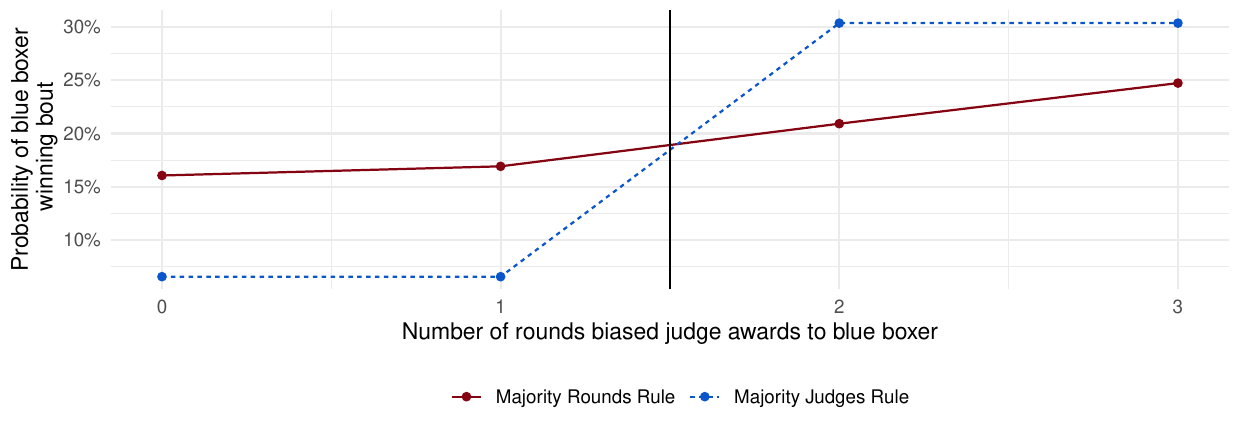}
	\caption{Probability of favoured boxer winning in one bout where the \textit{B}lue boxer won 1 round and the \textit{R}ed boxer won 2 (in the absence of noise) - Men's Olympic amateur boxing}
	\label{fig:curve6_menama}
 \end{figure}

\begin{figure}[ht!]
	\centering
		\includegraphics[width=0.85\textwidth]{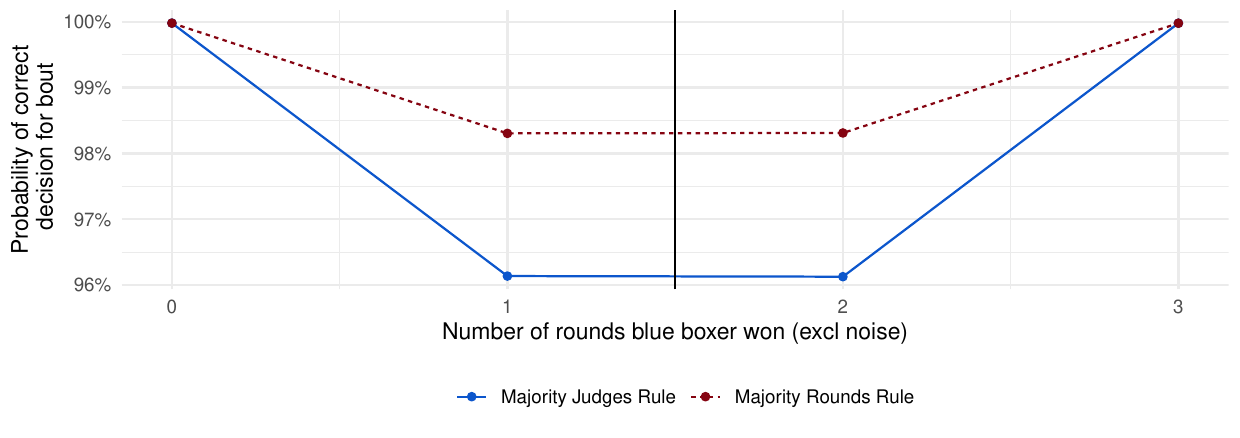}
	\caption{Probability of correct result for bout - Men's Olympic amateur boxing}
	\label{fig:outcomes_of_bouts_menama}
 \end{figure}

\begin{figure}[ht!]
	\centering
		\includegraphics[width=0.85\textwidth]{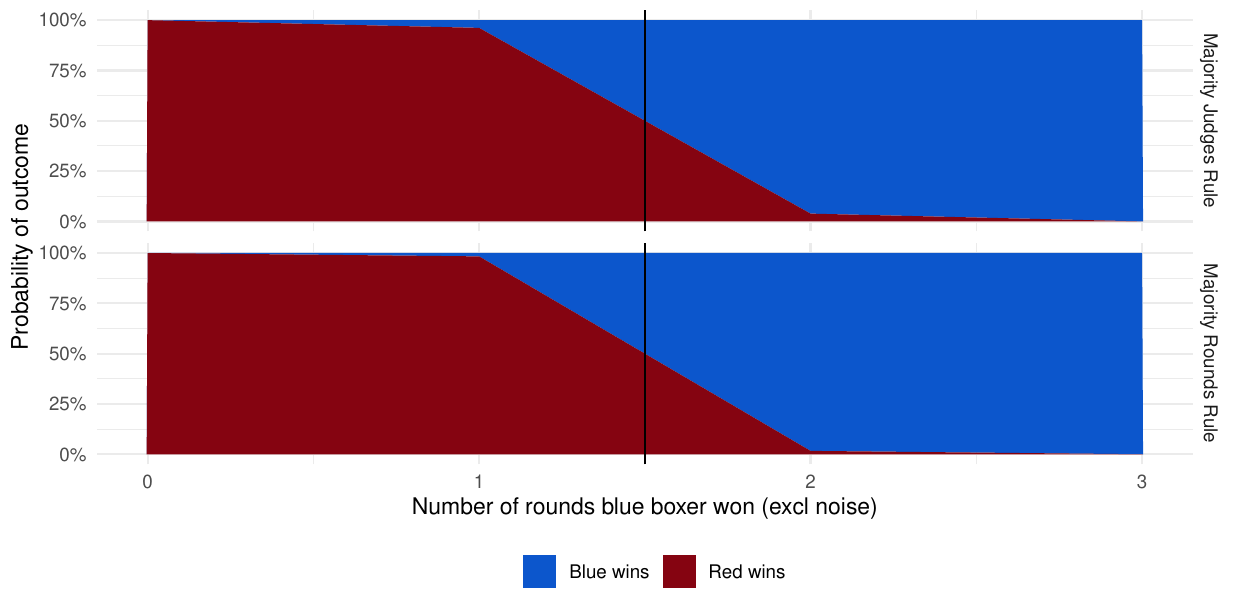}
	\caption{Probability of each outcome for bout - Men's Olympic amateur boxing}
	\label{fig:outcomes_of_bouts2_menama}
\end{figure}

\noindent In men's Olympic amateur boxing, there are generally 3 rounds and 5 judges. Figures \ref{fig:curve6}-\ref{fig:outcomes_of_bouts2} are reproduced below for men's Olympic amateur boxing as Figures~\ref{fig:curve6_menama}-\ref{fig:outcomes_of_bouts2_menama}.

  \vspace{-0.5cm}
  \vspace{-0.5cm}

\clearpage

\section{Women's Olympic amateur}\label{RepeatEverythingForWomensAmateur}
\renewcommand\thetable{\thesection\arabic{table}}
\renewcommand\thefigure{\thesection\arabic{figure}}
\renewcommand{\theequation}{\Alph{section}.\arabic{equation}}
\setcounter{equation}{0}
\setcounter{table}{0}
\setcounter{figure}{0}

\begin{figure}[ht!]
	\centering
 	\caption{Probability of favoured boxer winning in one bout where both boxers won 6 rounds (in the absence of noise)  - Women's Olympic amateur boxing}
		\includegraphics[width=0.85\textwidth]{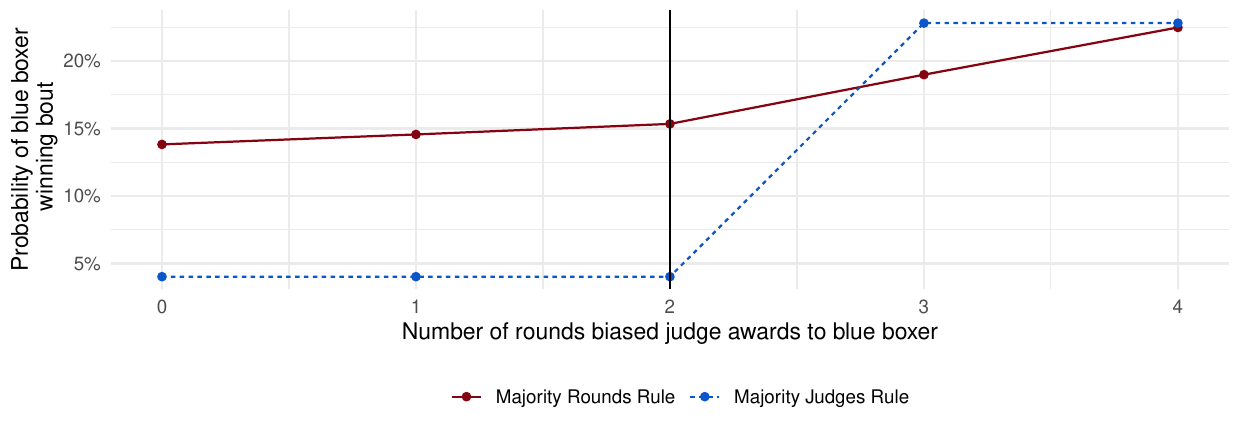}
	\label{fig:curve6_wama}
 \end{figure}

\begin{figure}[ht!]
	\centering
 	\caption{Probability of correct result for bout - Women's Olympic amateur boxing}
		\includegraphics[width=0.85\textwidth]{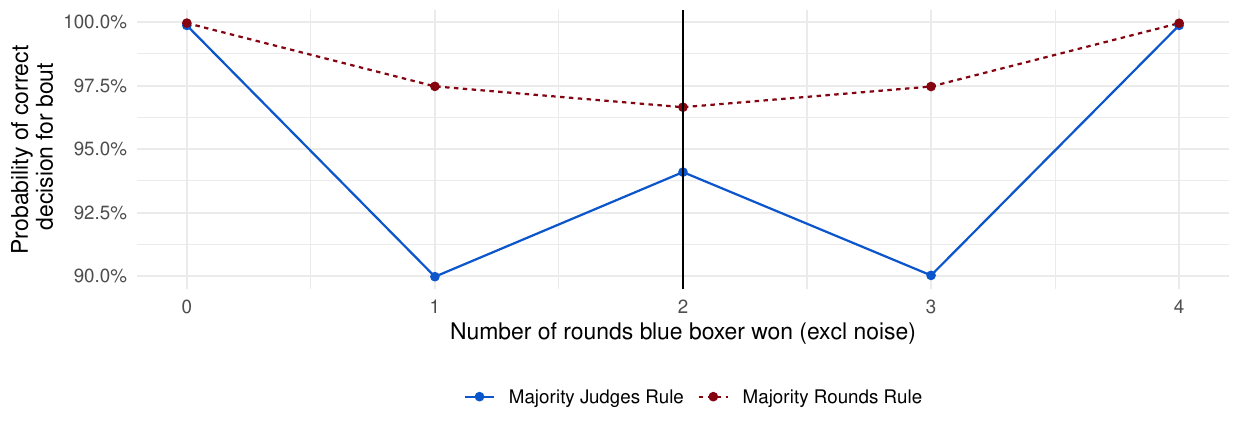}
	\label{fig:outcomes_of_bouts_wama}
 \end{figure}

\begin{figure}[ht!]
	\centering
 	\caption{Probability of each outcome for bout  - Women's Olympic amateur boxing}
		\includegraphics[width=0.85\textwidth]{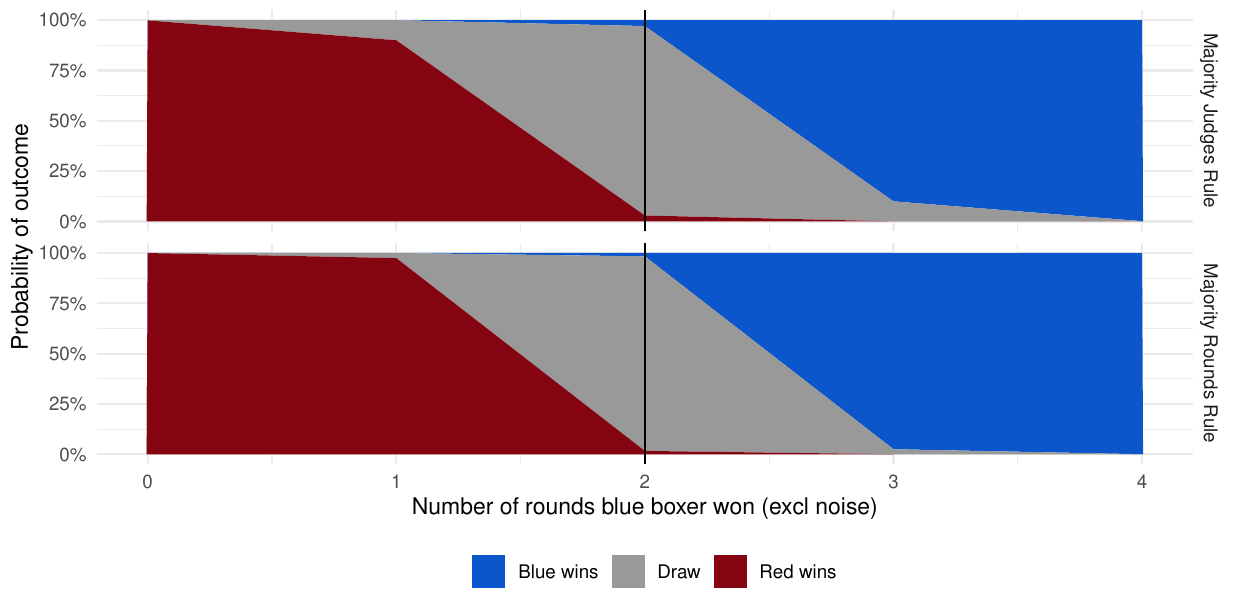}
	\label{fig:outcomes_of_bouts2_wama}
\end{figure}

\noindent In women's Olympic amateur boxing, there are generally 4 rounds and 5 judges. Figures \ref{fig:curve6}-\ref{fig:outcomes_of_bouts2} are reproduced below for women's Olympic amateur boxing as Figures \ref{fig:curve6_wama}-\ref{fig:outcomes_of_bouts2_wama}.
  \vspace{-0.5cm}
  \vspace{-0.5cm}

\clearpage

\end{appendices}
\end{document}